\documentclass[aps,preprint]{revtex4}
\newcommand{\be}{\begin{eqnarray}}
\newcommand{\ee}{\end{eqnarray}}

\begin{document}


\begin{titlepage}

\title{\bf Does Fluctuating Nonlinear Hydrodynamics Support an Ergodic-Nonergodic Transition?}

\author{Shankar P. Da$\mathrm{s}^1$ and Gene F. Mazenk$\mathrm{o}^2$\\
${\ }^1$ {\it School of Physical Sciences,
Jawaharlal Nehru University}\\
{\it New Delhi 110067, India}\\
${\ }^2$ {\it The James Franck Institute and the Department of Physics}\\
{\it The University of Chicago}\\
{\it Chicago, Illinois 60637, USA}}

\vspace*{1cm}

\begin{abstract}
Despite its appeal, real and simulated glass forming systems do not
undergo an ergodic-nonergodic  (ENE) transition. We reconsider
whether the fluctuating nonlinear hydrodynamics (FNH) model for this
system, introduced by us in 1986, supports an ENE transition. Using
nonperturbative arguments, with no reference to the hydrodynamic
regime, we show that the FNH model does not support an ENE
transition.  Our results  support the findings in the original
paper. Assertions in the literature questioning the validity of the
original work are shown to be in error.
\end{abstract}

\vspace*{1cm}


\maketitle
\end{titlepage}

\section{Introduction}

It is appealing to associate the vitrification of the liquid into a
frozen glassy state as a ergodic nonergodic (ENE) transition.
Unfortunately there is strong evidence against the ENE transition
scenario in physical and numerical experiments. This is in agreement
with the results we found twenty years ago in Ref. \cite{DM} (
hereafter mentioned as DM) when we introduced the model of
fluctuating nonlinear hydrodynamics (FNH). We present here a
nonperturbative analysis of the FNH model and the possibility of an
ENE transition. In the end our results here agree with those
in Ref. \cite{DM}.  There is no sharp ENE transition in the FNH
model. Recent reservations \cite{ABL} concerning our results in are
shown to be unfounded. We also address some
misrepresentations\cite{CR} of our work.

In the theory of Classical Liquids, a new approach to studying the
complex behavior of the supercooled state started with the
introduction of the self consistent mode-coupling theory
(MCT)\cite{rmp,adv}. The model referred to here is based on a
nonlinear feedback mechanism due to the coupling of the slowly
decaying density fluctuations in the supercooled liquid. The feed
back effects at metastable densities strongly enhance the transport
properties of the liquid. In the simple version proposed initially
\cite{leth,beng,prl84} a sharp ergodic to non-ergodic (ENE)
transition of the liquid into a glassy phase was predicted. This
transition occurs  at a critical density ( or at the corresponding
values of other controlling thermodynamic parameters) beyond which
the density auto correlation function freezes at a nonzero value
over long times. Soon afterward it was demonstrated that this sharp
ENE transition is \cite{DM}  rounded. The absence of a sharp ENE
transition in the supercooled liquids was supported by work
\cite{goetze87,dufty} using  similar theoretical models. Two recent works
\cite{ABL,CR} has called these conclusions into question.  The
purpose of the present paper is to show that our previous analysis
withstands careful scrutiny and
to reassert that the results of ref. \cite{DM} are correct and
captures the right phenomena for the removal of the ENE transition.

We organize this paper as follows. In the next section we briefly
introduce the FNH model.  This is followed by an analysis of whether
this model supports an ENE transition.   In section III we compare
our findings here to those in DM.  Next we comment on the works
which question the conclusions in DM. We end the
paper with a short discussion.\\

\section{The Fluctuating Nonlinear Hydrodyanmic Model}

In Ref. \cite{DM} a model for the long time relaxation behavior of
the supercooled liquid was constructed using  fluctuating nonlinear
hydrodynamics. The dynamics of collective modes in the liquid was
formulated with {\em nonlinear} Langevin equations involving bare
transport coefficients. These nonlinear stochastic equations  for
the time evolution of the conserved densities are plausible
generalizations of the macroscopic hydrodynamic laws. The set of
collective variables $\{\psi_i\}$ for the liquid we considered
consists of mass and momentum densities $\{\rho({\bf r},t), {\bf
g}({\bf r},t)\}$. The construction of the equations of motion
\cite{ma_mazenko} for the slow variables involve a driving free
energy functional $F$ which is expressed in terms of the
hydrodynamic fields, {\em i.e.}, $\rho$ and ${\bf g}$. The
corresponding equilibrium distribution for the system is
$\exp(-{\beta}F)$. The free energy functional $F$ is separated in
two parts, $F=F_K[{\bf g},\rho]+F_U [\rho]$. The dependence of $F$
on ${\bf g}$ is entirely in the kinetic part $F_K$ in the form
\cite{langer} constrained by galilean invariance:
\begin{equation}
\label{kin-f} F_K [{\bf g},\rho] ~=~ \int d{\bf x} \frac{g^2 ({\bf
x})}{2\rho ({\bf x})}~.
\end{equation}
\noindent The potential part $F_U$ is treated as a functional of the
density only. The density $\rho$ follows the continuity equation

\begin{equation}
\label{cont} {{\partial \rho} \over {\partial t}} +{\bf \nabla}.{\bf
g} =0,
\end{equation}

\noindent having the flux as the momentum density ${\bf g}$ which
itself is a conserved property. The nonlinear equation for the
momentum current density $g_i$ is a generalized form of the
Navier-Stokes equation\cite{DM},

\begin{equation}
\label{g-DM} {{\partial g_i} \over {\partial t}} = - \sum_j \nabla_j
[\frac{g_i g_j}{\rho}] - \rho {\nabla}_i \frac{\delta F_U}{\delta
\rho} - \sum_j L^o_{ij} \frac{g_j}{\rho} + \theta_i~.
\end{equation}

\noindent The noise $\theta_i$ is assumed to be Gaussian following
the fluctuation dissipation relation to the bare damping  matrix
$L^o_{ij}$. For compressible liquids, the $1/\rho$ non-linearity
appear in two terms in the generalized Navier-Stokes equation. These
are respectively  the convective term coupling two flow fields and
the dissipative term involving the bare viscosity of the liquid. The
appearance of this non linearity in the hydrodynamic equations is
formally avoided in Ref.\cite{DM} by introducing the local velocity
field ${\bf V}({\bf x},t)$,
\begin{equation}
\label{compr} {\bf g} ({\bf x},t) = \rho ({\bf x},t) {\bf V} ({\bf
x},t).
\end{equation}

\noindent The set of fluctuating variables in terms of which the
renormalized field theory is constructed in our analysis therefore
consists of the set $\psi_i \equiv \{\rho,{\bf g},{\bf V} \}$.

The consequences of the nonlinearities in the equations of motion,
{\em i.e.}, renormalization of bare transport coefficients, are
obtained using graphical methods of field theory \cite{martin}. The
correlation of the hydrodynamic fields involve averages defined in
terms of the action ${\cal A}$ which is a functional of the field
variables $\{\psi_i\}$ and the corresponding conjugate hatted fields
$\{\hat{\psi}_i\}$ introduced in the MSR formalism. Using the
equations of motions (\ref{cont}) and (\ref{g-DM}) respectively for
$\rho$ and ${\bf g}$ the action functional is obtained as \cite{DM},

\begin{eqnarray}
\label{action-MSR} {\cal A} &=& \int{dt}\int{d{\bf x}}\left \{
\sum_{i,j} \hat{g}_i\beta^{-1}L^o_{ij}\hat{g}_j
 +i\sum_{i} \hat{g}_i \left [ \frac{\partial g_i}{\partial t} +
\rho \nabla_i \frac{\delta F_u}{\delta \rho}+ \sum_j \nabla_j (\rho
V_i V_j) - \sum_j L^o_{ij} V_j \right ] \right.
\nonumber \\
~~~.
&+& \left. i \hat{\rho} \left[\frac{\partial \rho}{\partial t}+ {\bf
\nabla}.{\bf g} \right ] +i \sum_i \hat{V}_i [ g_i-\rho V_i ] \right
\} ~,
\end{eqnarray}

\noindent The theory is developed in terms of the correlation
functions,

\begin{equation} \label{g-function}
 G_{\alpha\beta}(12)=
\langle\psi_{\beta} (2)\psi_{\alpha}(1)\rangle \end{equation}
\noindent  and the response functions,
 \begin{equation}
 \label{r-function}
G_{\alpha\hat{\beta}}(12)= \langle\hat{\psi}_{\beta}
(2)\psi_{\alpha}(1)\rangle ~~~.
\end{equation}
The averages here are functional integrals over all the fields weighted
by $e^{-A}$.

\noindent The nonlinearities in the equations of motion (\ref{g-DM})
and (\ref{compr}) give rise non-gaussian terms in the action
(\ref{action-MSR}) involving products of three or more field
variables. The role of the non gaussian parts of the action ${\cal
A}$ on the correlation functions are quantified in terms of the self
energy matrix which show up in the equation satisfied by the
response functions and that satisfied by the correlation functions.
We begin with the response functions which satisfy:
\begin{equation} \label{dyson-resp}
\left[(G_{0}^{-1})_{\hat{\alpha}\mu}(13)
-\Sigma_{\hat{\alpha}\mu}(13) \right]G_{\mu\hat{\beta}}(32)= \delta
(12)\delta_{\hat{\alpha}\hat{\beta}},
\end{equation}

\noindent with self energies $\Sigma_{\hat{\alpha}\mu}$ which can be
expressed in perturbation theory in terms of the two-point
correlation and response functions. Using the explicit polynomial form of the
action (\ref{action-MSR}), the response functions are expressed in
the general form,

\begin{equation} \label{resp-fun}
 G_{\alpha
\hat{\mu}}=\frac{N_{\alpha\hat{\mu}}}{D} \label{eq:12}
\end{equation}

\noindent where the matrix $N$ is given in table I
and the determinant $D$ in the denominator is given by

\begin{equation}
\label{detrmin}D=\rho_\mathrm{L}(\omega^{2}-q^{2}c^{2})+iL (\omega
+iq^{2}\gamma) ~~~.
\end{equation}

\noindent The various quantities are defined such that
$\rho_\mathrm{L}$ ,$c^{2}$ and $L$ are identified as the
corresponding renormalized quantities respectively for the bare
density $\rho_0$, speed of sound squared $c_0^2$ and longitudinal
viscosity $L_0$. We have in terms of single-hatted or response
self-energies:,

\begin{eqnarray}
\label{cut-self}
\ \rho_\mathrm{L} &=& \rho_{0}-i\Sigma_{\hat{V}V}\\
\label{visc-ren}
\ L&=& L_0 +i\Sigma_{\hat{g}V}\\
\label{sound-ren} qc^{2}&=& qc_{0}^{2}+\Sigma_{\hat{g}\rho}
\end{eqnarray}

\noindent and $\gamma$ is defined in terms of the self energy
element $\Sigma_{\hat{V}\rho} \equiv q\gamma$. One can also
show that the correlation functions of the
physical un-hatted field variables are given by,

\begin{equation}
 \label{eq:84} G_{\alpha\beta}=-\sum_{\mu\nu
} G_{\alpha ,\hat{\mu}}C_{\hat{\mu}\hat{\nu}} G_{\hat{\nu}\beta}
\end{equation}

\noindent where Greek letter subscripts take values $\rho ,{\bf g},
{\bf V}$, and the self energy matrix $C_{\hat{\mu}\hat{\nu}}$ is given by,

\begin{equation}\label{c-matrix}
 C_{\hat{\mu}\hat{\nu}}= 2\beta^{-1}L_{0}\delta_{\hat{\mu}\hat{\nu}}
\delta_{\hat{\mu},\hat{g}} -\Sigma_{\hat{\mu}\hat{\nu}} ~~~.
\end{equation}

\noindent The double-hatted self-energies
$\Sigma_{\hat{\mu}\hat{\nu}}$ vanish if either index corresponds to
the density. This model does not have
a complete set of FDR linearly relating
correlation and response functions.

However, using the time translational
invariance properties of the action (\ref{action-MSR}), we obtained in DM
the following fluctuation dissipation relation between correlation
and response functions involving the field $g$ in the form :

\begin{equation}
\label{DM-gFDT} G_{V_i\alpha} (q,\omega) = -2\beta^{-1} {\rm Im}
G_{\hat{g}_i\alpha} (q,\omega) \label{eq:17}
\end{equation}

\noindent where $\alpha$ indicates any of the fields $\{\rho,g,V\}$.

\section{Ergodic-Nonergodic transition and FNH}

Does this model have an ENE transition? To answer this question we
first pose the conditions for such a transition.  Suppose, due to a
nonlinear feedback mechanism, the self-energy
$\Sigma_{\hat{g}\hat{g}}$ blows up at small frequencies:

\begin{equation} \Sigma_{\hat{g}\hat{g}} =-A\delta (\omega ) ~~~. \label{eq:22}
\end{equation}

\noindent This is presumed to result from a persistent time
dependence of the density correlation function. This hypothesis is
motivated by the one-loop contribution and the physics of the
viscosity blowing up as one enters the glass. Is this assumption
compatible with the set of Dyson equations? What we mean by a
nonergodic phase is that $G_{\rho\rho}$ shows a $\delta$-function
peak at zero frequency. Putting Eq.(\ref{eq:22}) back into
Eq.(\ref{eq:84}) we obtain a $\delta (\omega )$ peak in
$G_{\rho\rho}$ as long as the response function $G_{\rho\hat{g}}$ is
not zero in the $\omega\rightarrow 0$ limit.  We assume, with no reason to
expect otherwise, that the $\omega\rightarrow 0$ limits of
$\rho_\mathrm{L}$, $\gamma$, $c^{2}$ and $L$ are nonzero.  With
these assumptions $D$ is not infinite in the low frequency limit and
$G_{\rho\hat{g}}$, and $G^{\mathrm{L}}_{Vg}$ are nonzero in the low
frequency limit. Then from Eq.(\ref{eq:84}) we find that
$G_{\rho\rho}$, $G_{\rho V}$, and $G_{VV}$ show a $\delta (\omega )$
component. Since $G_{g\hat{g}}$ vanishes as $\omega\rightarrow 0$ as
long as $D(\omega =0)\neq 0$, the correlation functions involving a
momentum density index do not show a $\delta$-function peak at zero
frequency. So it is necessary for an ENE transition that
$G_{\rho\hat{g}}$ not vanish as $\omega\rightarrow 0$. This requires
that $\rho_{\mathrm{L}}$ goes to a nonzero value in the zero
frequency limit and the determinant $D$ not blow up as
$\omega\rightarrow 0$.

If, as expected, the self-energy contibution $\gamma (\omega =0)\neq
0$ then the correlation functions $G_{\rho V}$, and  $G_{VV}$ show a
$\delta (\omega )$ component. Now we apply  the FDT (\ref{DM-gFDT}).
Since $G_{V\rho}$ and $G_{VV}$ blow up, it then follows from the FDT
that the imaginary parts of the response functions $G_{\hat{g}\rho}$
and $G_{\hat{g}V}$ also blow up. However we also require
simultaneously that $D^{*}D$ is bounded, and imaginary parts of both
$\rho_{\mathrm{L}}qD^{*}$ and $(\omega +iq^{2}\gamma)D^{*}$ diverge.
But since both $D'$ and $D''$ denoting the real and imaginary parts
of $D$ are bounded so $\rho_\mathrm{L}$ and $\gamma$ must diverge.
However, if these quantities blow up then from (\ref{detrmin}) it
follows that $D$ must also blow up and we have a contradiction. The
obvious conclusion is that the original assumption of a nonergodic
phase is not supported in the model. The key self-energy
contribution is $\gamma$.  If for some reason this quantity vanishes
at zero frequency then $G_{\rho V}$, and $G_{VV}$ vanish as $\omega$
goes to zero.  Then $G_{\rho V}$, and $G_{VV}$ do not show a $\delta
(\omega )$ component and one does not have the constraints on
$\rho_\mathrm{L}$, $\gamma$, and $D$. In this case one may have an
ENE transition in this model.

\section{Relation to DM Results}
The argument we give in the hydrodynamic regime in Ref. \cite{DM} is
completely consistent with the
results presented above.
The simplest way of understanding the argument in the previous
section is to look at the response function
\begin{equation}
G_{\rho\hat{\rho}}=\frac{\omega\rho_\mathrm{L} +iL}
{\rho_\mathrm{L}(\omega^{2}-q^{2}c^{2})+iL(\omega +iq^{2}\gamma)}
~~~. \label{dens-resp} \end{equation}

\noindent The renormalization of the longitudinal viscosity $L$ is
computed, see Eq.(\ref{visc-ren}) in terms of the longitudinal part
$\Sigma^L_{\hat{g}V}$ of the corresponding self-energy matrix
$\Sigma_{\hat{g}_iV_j}$ of the isotropic liquid,

\begin{equation}
\label{Gam-hy-c} L(q,z) = L_0+\frac{\beta}{2}
\Sigma^\mathrm{L}_{\hat{g}V}(q,z).
\end{equation}
\noindent If we ignore the self energy $\Sigma_{\hat{V}\rho}$, the
expression (\ref{dens-resp}) is identical to the conventional
expression for the density correlation function  with the
generalized memory function or the renormalized transport
coefficient $L(q,z)$. The dependence of  $G_{\rho\hat{\rho}}$ on the
self energy $\Sigma_{\hat{V}\rho}$ in the renormalized theory is a
consequence of the non-linear term involving the $\hat{V}$ field in
the MSR action (\ref{action-MSR}) and is originating from the
nonlinear constraint (\ref{compr}) introduced to deal with the
$1/\rho$ nonlinearity in the hydrodynamic equations.
Analyzing the expression (\ref{eq:84}) for the correlation functions
and the FDT relation  (\ref{DM-gFDT}) we obtain in the hydrodynamic
limit the following nonperturbative relation between the two types of
 self energies contributing alternatively to the renormalization of the
longitudinal  viscosity,

\begin{equation}
\label{visc-renm} \gamma_{\hat{g}\hat{g}} (0,0)= 2\beta^{-1} \left [
\gamma^\prime_{\hat{g}V}(0,0)+ \lim_{\omega{\rightarrow}0} \left(
\gamma^{\prime\prime}_{\rho\hat{g}}(0,\omega)/\omega \right )
 \right ]
\end{equation}

\noindent where we have used in the above following definitions, in
the isotropic limit, $\Sigma^\mathrm{L}_{\hat{g}\hat{g}} \sim
 - q^2\gamma_{\hat{g}\hat{g}}$, $\Sigma^L_{\hat{g}V} \sim -
iq^2\gamma_{\hat{g}V}$, and $\Sigma_{\rho\hat{g}} \sim
q\gamma_{\rho\hat{g}}$.
The relation (\ref{visc-renm}) which is
obtained from the FDT relation (\ref{DM-gFDT}) only, implies that
both the self energies $\Sigma_{\hat{g}\hat{g}}$ and
$\Sigma_{\hat{g}V}$ have the same diverging contribution in the low
frequency limit.  In the simplified model it is this contribution in
terms of density correlation function which constitute the feed back
mechanism of MCT and leads to the dynamic transition beyond a
critical density. The singular contribution to the renormalized
transport coefficient $L$ in (\ref{dens-resp}) is now obtained in
terms of the self energy $\Sigma_{\hat{g}V}$. As a consequence of
(\ref{visc-renm}) it also follows that the response function
$G{\rho\hat{\rho}}$ is equal to the corresponding density
correlation function $G_{\rho\rho}$  in the hydrodynamic limit. It
is important to note here that ( contrary to the assertion in Ref. \cite{ABL}) this
relation is not forced by us, rather it follows as a natural
consequence of the relations (\ref{visc-renm})  linking the response
to correlation self energies.

The asymptotic behavior of the density correlation function is
inferred from  $G{\rho\hat{\rho}}$. The
denominator of  (\ref{dens-resp}) for the
response functions contain the self energy $\Sigma_{\hat{V}\rho}$
which in this case is crucial for the long time dynamics and
understanding how the ENE transition is cutoff. The density
correlation function ( in the small $q$, $\omega$ limit) only freeze
due to the feed back mechanism {\em if} the self energy matrix
elements $\Sigma_{\hat{V}\rho}$ is zero - a result obtained in the
earlier section. In this regard it is useful to note that for
the $\omega \rightarrow 0$ limit the quantity
$L(\omega+i\gamma q^2)$ in $D$ does not diverge
 even when $L \sim 1/\omega $ is getting large, since
$L\gamma{q^2}$ remains finite in the non hydrodynamic regime
$\omega \sim q^2$.
To leading order in wave numbers
$q{\Sigma_{\hat{V}\rho}}(q,0) \equiv q^2{\gamma}$, is expressed in
terms of the self energy $\Sigma^\mathrm{L}_{\hat{V}\hat{V}}$ using
the nonperturbative relation

\begin{equation}
\label{cut-hyd}
\gamma_{\hat{V}\hat{V}}(0,0)=
\frac{2\rho{\beta^{-1}}}{c^2} \gamma^\prime_{\rho\hat{V}}(0,0),
\end{equation}

\noindent where $c$ is the sound speed introduced in
 (\ref{sound-ren}).
Note that the relation (\ref{cut-hyd}) is also {\em obtained from
the same fluctuation-dissipation relation (\ref{DM-gFDT})}.

\section{ABL and CR}
We now address the criticisms made in Ref. \cite{ABL} on our work.
ABL imply that we misapplied the FDT relating $G_{\rho\rho}$ and
$G_{\rho\hat{\rho}}$ in the hydrodynamical limit. These authors
offer that we assumed a linear FDT from the beginning.
DM clearly discusses the consequences of not having a complete set
of FD relations.
 On the other hand the introduction of the
 $\theta = \delta F / \delta \rho$ field by ABL for obtaining a FDT in the
 linear form has not yet been shown to be useful.
The nonlinear contribution in $\theta$ comes
 from the part $\delta F_K / \delta \rho$, which actually gives rise
 \cite{DM} to the term $\nabla_j (g_i g_j/\rho)$ in the generalized
 Navier-Stokes equation. The latter is essentially the $1/\rho$ nonlinearity
 which we address in our model through the introduction of the
 variable $V$.
 In this regard we believe that the importance of linear FDT in the MSR formulation
  has been overemphasized by ABL.
 Indeed in the absense of a linear
 FDT the response functions lose their physical meaning and become mere
 computational tools. From a  physical point of view however what is important is that the correlation functions are time invariant which is maintained as can be directly seen from the above equation (\ref{eq:84}).

In Ref. \cite{CR} the cut off mechanism of Ref. \cite{DM} has been
questioned by treating the highly nonlinear model described above using
a rather naive approach.
CR basically make some phenomenological manipulations on a Newtonian
dynamics model \cite{z1}, ending with a memory function description they
claim, {\em without proof}, is related to our model. All subsequent
discussion of our work made by these authors are based on
this claim.
Our model, as shown on
examination of table I, satisfies at all stages
the density conservation law.
The memory function proposed in CR to represent our work, Eq.(5)
there, does not satisfy this conservation law.  Therefore the
analysis of CR does not apply to the model we studied. None of their
conclusions concerning our work have any validity.
CR concede that
there is no error in our calculation, rather they offer vaguely that
our model itself is the problem!
Our model, as shown on
examination of table I, satisfies at all stages
the density conservation law.
The source of their error appears to be the naive assumption that this model can
be represented in terms of a single memory function \cite{kawasaki-94}.
This work represents a fundamental misunderstanding of the problem.

Though the authors of both papers, ABL and CR,  seem to agree that finally the
ENE transition does not survive
they disagree with our analysis of the problem. The
arguments put forward in Ref. \cite{CR} to rediscover that the
transition is finally cutoff are rather vague and of descriptive
nature. These authors only seem to conjecture that the
transition will be cutoff nonperturbatively citing other recent
works \cite{mayer}.

\section{Discussion}

The basic feedback mechanism of MCT is a consequence of simple
quadratic nonlinearities in density fluctuations ( arising from
purely dynamic origin) that is present in the pressure term of the
generalized Navier-Stokes equation. The ergodicity restoring
mechanism goes beyond this. The description in terms of coupling to
currents is a physically appealing way of explaining the nature of
the FNH equations ( expressed in a form which can be sensibly
related to the hydrodynamics of liquids). It is in fact the full
implications of the density nonlinearities in the dynamics that cuts
off the sharp transition to nonergodicity.
This is also reflected in
the fact that the basic conclusions of Ref. \cite{DM} follow even if
the relevant nonlinearity is considered in a different manner.
In fact by formulating the model\cite{yeo-stat} only in terms of the fields
$\{\rho,{\bf g}\}$ the same conclusions implying the absence of the
dynamic transition is reached as in Ref. \cite{DM}. The $1/\rho$
nonlinearity mentioned above is treated here in terms of a series of
density nonlinearites. The self energy matrix elements
$\Sigma_{\hat{g}V}$ and $\Sigma_{\hat{V}\rho}$ are absent from the theoretical formulation in this case and the cutoff kernel is obtained here from a different
self energy element $\Sigma_{\hat{g}\rho}$.

Twenty years ago we had predicted that the
feedback effects from mode-coupling of density fluctuations, when
properly analyzed keeping consistency with concepts of basic
hydrodynamics, results in a qualitative crossover in the dynamics.
We presented here a selfcontained nonperturbative proof that FNH
does not support an ENE transition.  This new analysis is completely
compatible with the results of DM, simulations and experiment.


\section*{Acknowledgement}
SPD acknowledges CSIR, India for financial support.

\begin{center}
\begin{table}
\label{resp-table}
\begin{tabular}{|c|ccc|}
\hline \multicolumn{1}{|c|}{}
&\multicolumn{1}{c}{$\hat{\rho}$}&\multicolumn{1}{c}{$\hat{g}$}&\multicolumn{1}{c|}{$\hat{V}$} \\
\hline
 $ \rho $ &~~~~~~$\omega\rho_{\mathrm{L}} +iL $ & ~~~~~$\rho_{\mathrm{L}}q$~~~~~&~~~~~$Lq$~~~~~~\\
  $g$ &~~~~~~$ q(\rho_{\mathrm{L}}c^{2}+L\gamma)$&~~~~~~$\rho_{\mathrm{L}}\omega$~~~~~~&~~~~~~$L\omega$~~~~~~\\
$ V$ &~~~~~~~$ q(c^{2}+i\omega\gamma )$& $\omega +iq^{2}\gamma $
&~~~~~~$i(\omega^{2}-q^{2}c^{2})$~~~~~~~ \\
\hline
\end{tabular}
\caption{The matrix of the coefficients $N_{\alpha\hat{\beta}}$ in
the numerator on the RHS of eqn. (\ref{resp-fun}) for the response
functions}
\end{table}
\end{center}

\end{document}